\documentclass[onecolumn,amsmath,showpacs,nofootinbib,11pt]{revtex4-2}
\usepackage{graphicx}
\usepackage{dcolumn}
\usepackage{bm}
\usepackage{color} 
\usepackage{slashed}
\usepackage{amsfonts}
\begin{document}
\newcommand{\hs}{\hspace*{0.5cm}}
\newcommand{\vs}{\vspace*{0.5cm}}
\newcommand{\be}{\begin{equation}}
\newcommand{\ee}{\end{equation}}
\newcommand{\bea}{\begin{eqnarray}}
\newcommand{\eea}{\end{eqnarray}}
\newcommand{\ben}{\begin{enumerate}}
\newcommand{\een}{\end{enumerate}}
\newcommand{\bde}{\begin{widetext}}
\newcommand{\ede}{\end{widetext}}
\newcommand{\nn}{\nonumber}
\newcommand{\crn}{\nonumber \\}
\newcommand{\Tr}{\mathrm{Tr}}
\newcommand{\non}{\nonumber}
\newcommand{\noi}{\noindent}
\newcommand{\al}{\alpha}
\newcommand{\la}{\lambda}
\newcommand{\bet}{\beta}
\newcommand{\ga}{\gamma}
\newcommand{\va}{\varphi}
\newcommand{\om}{\omega}
\newcommand{\pa}{\partial}
\newcommand{\+}{\dagger}
\newcommand{\fr}{\frac}
\newcommand{\bc}{\begin{center}}
\newcommand{\ec}{\end{center}}
\newcommand{\Ga}{\Gamma}
\newcommand{\de}{\delta}
\newcommand{\De}{\Delta}
\newcommand{\ep}{\epsilon}
\newcommand{\varep}{\varepsilon}
\newcommand{\ka}{\kappa}
\newcommand{\La}{\Lambda}
\newcommand{\si}{\sigma}
\newcommand{\Si}{\Sigma}
\newcommand{\ta}{\tau}
\newcommand{\up}{\upsilon}
\newcommand{\Up}{\Upsilon}
\newcommand{\ze}{\zeta}
\newcommand{\ps}{\psi}
\newcommand{\Ps}{\Psi}
\newcommand{\ph}{\phi}
\newcommand{\vph}{\varphi}
\newcommand{\Ph}{\Phi}
\newcommand{\Om}{\Omega}
\newcommand{\AdrHEPC}{Phenikaa Institute for Advanced Study, Phenikaa University, Yen Nghia, Ha Dong, Hanoi 100000, Vietnam}
\newcommand{\AdrIOP}{Institute of Physics, Vietnam Academy of Science and Technology, 10 Dao Tan, Ba Dinh, Hanoi 100000, Vietnam}
\newcommand{\AdrUETI}{Faculty of Applied Sciences, University of Economics - Technology for Industries, 456 Minh Khai, Hai Ba Trung, Hanoi, Vietnam}

\title{Dark symmetry implication for right-handed neutrinos} 

\author{Phung Van Dong}
\email{dong.phungvan@phenikaa-uni.edu.vn}
\affiliation{\AdrHEPC} 
\author{Duong Van Loi}
\email{loi.duongvan@phenikaa-uni.edu.vn (corresponding author)}
\affiliation{\AdrHEPC} 
\author{Do Thi Huong}
\email{dthuong@iop.vast.vn}
\affiliation{\AdrIOP} 
\author{Nguyen Tuan Duy}
\email{ntduy@iop.vast.vn}
\affiliation{\AdrIOP} 
\author{Dang Van Soa}
\email{soadangvan@gmail.com}
\affiliation{\AdrUETI} 

\date{\today}

\begin{abstract}

We argue that the long-standing issues of neutrino mass and dark matter can be manifestly solved in a dark gauge symmetry $U(1)_D$ that transforms nontrivially only for three right-handed neutrinos $\nu_{1,2,3R}$---the counterparts of known left-handed neutrinos. This theory assigns $\nu_{1,2,3R}$ dark charge to be $D=0$, $-1$, and $+1$, respectively, in order for anomaly cancelation. Additionally, it imposes an inert Higgs doublet $\eta$ and two Higgs singlets $\xi,\phi$ with dark charge $D=+1$, $-1$, and $+2$, respectively. That said, the dark symmetry is broken by $\phi$ (by two units) down to a dark parity $P_D=(-1)^D$, for which $\nu_{2,3R}$ and $\eta,\xi$ are odd, whereas all other fields are even due to $D=0$. The lightest of these odd fields is stabilized by $P_D$, responsible for dark matter. Neutrino masses are generated by a scotoseesaw scheme, in which the seesaw part is mediated by $\nu_{1R}$, while the scotogenic part is mediated by $\nu_{2,3R}$, for which the hierarchy of atmospheric and solar neutrino mass splittings is explained.         

\end{abstract}

\maketitle

\section{Introduction}

Neutrino mass \cite{kajita,mcdonald} and dark matter \cite{bertone,arcadi} are among the leading questions in science, which prove that the standard model must be extended. If right-handed neutrinos---respective counterparts of left-handed neutrinos---are supposed, the seesaw mechanism takes place, inducing small neutrino masses as suppressed by the right-handed neutrino mass scale \cite{Minkowski,Yanagida,Gell-Mann,Mohapatra,valle}. Unfortunately, the seesaw mechanism in itself does not address dark matter stability. 

If right-handed neutrinos are odd under a $Z_2$, while standard model particles are even under this group, the seesaw mechanism is suppressed. Left-handed neutrinos can now couple to right-handed neutrinos through an imposed inert Higgs doublet that is odd under $Z_2$ too. In this way, neutrino masses are radiatively induced by one-loop diagrams contributed by right-handed neutrinos and inert Higgs fields, called scotogenic mechanism \cite{tao,ma}. This approach is motivated since the lightest of these odd fields either a right-handed neutrino or a neutral inert Higgs field is stabilized by $Z_2$ responsible for dark matter. However, the natural origin of $Z_2$ is a mystery. 

Such discrete symmetry, i.e. $Z_2$, would arise from a more fundamental gauge symmetry, as observed by Krauss and Wilczek long ago \cite{kw}. The simplest of which is perhaps a continuous abelian charge $U(1)_D$, which transforms nontrivially only for right-handed neutrinos, similarly to its residual $Z_2$. That said, label the right-handed neutrinos to be $\nu_{1,2,3R}$ that possess dark charge values as $D_{1,2,3}$, respectively, whereas all standard model fields are neutral, i.e. $D=0$. Nontrivial anomaly cancellation conditions according to $[\mathrm{Gravity}]^2 U(1)_D$ and $[U(1)_D]^3$ demand $D_1+D_2+D_3=0$ and $D^3_1+D^3_2+D^3_3=0$, respectively. They yield a unique solution, $D_1=0$, $D_2=-1$, and $D_3=+1$, given that we both indistinguish $\nu_{1,2,3R}$ states and normalize a nonzero charge to unity, without loss of generality, in agreement to \cite{dongloi}. 

Since $\nu_{1,2,3 R}$ dark charges are not universal, the above setup generally requires a plenty of scalar fields, say two extra Higgs doublets and one extra Higgs singlet~\cite{dongloi}, in order for neutrino mass generation. Additionally, one-loop neutrino mass generation comes from various diagram kinds, which are not physically distinct and complicated. This work argues that a minimal scalar content for this kind of model is available, for which the neutrinos obtain a mass via minimal scotoseesaw scheme, opposite to that in \cite{dongloi}. 

Let us attract the reader's attention to previous developments of scotoseesaw mechanism, which is implied by our setup \cite{kubo,valless,valless1,valless2,extrass3,extrass4}.\footnote{It is noted that a gauged $B-L$ symmetry that assigns $B-L=-4,-4,+5$ to $\nu_{1,2,3R}$ respectively also hints a scotoseesaw scheme \cite{valless5}. Further, the scotogenic part in such $B-L$ model may alternatively come from \cite{dongsg}.} The novelty of the scotoseesaw is that neutrino mass comes from both tree-level contribution by canonical seesaw for one family and loop-level contribution by scotogenic scheme for another family. Hereby, the scotoseesaw is well-motivated as addressing experimental hierarchical masses concerning the solar and atmospheric neutrino oscillations, while it still provides a dark matter candidate.     

The rest of this work is organized as follows. In Sec. \ref{model}, we propose the model, specifying dark symmetry, particle content, Lagrangian, symmetry breaking, dark parity, and physical scalar spectrum. In Sec. \ref{neutrino}, we investigate neutrino mass generation. In Sec. \ref{darkmatter}, we identify dark matter candidates and compute dark matter observables. In Sec. \ref{bound}, we present new physics signals and constraints. In Sec. \ref{flvg2}, we discuss charged lepton flavor violation and implication of the muon $g-2$ anomaly. We summarize our results and make concluding remarks in Sec. \ref{concl}.   

\section{\label{model}The model}

Add a dark symmetry $U(1)_D$ and three right-handed neutrinos $\nu_{1,2,3R}$ to the standard model. Assign $D$-charge for $\nu_{1,2,3R}$ to be $0$, $-1$, and $+1$, respectively, whereas all the standard model fields have $D=0$. Introduce an inert Higgs doublet $\eta$ and two Higgs singlets $\xi,\phi$ with $D$-charge $+1$, $-1$, and $+2$, respectively. The particle content of the model under the full gauge symmetry is collected in Table \ref{tab1}, where $P_D=(-1)^D$ is a residual dark parity of $U(1)_D$, specified later.  

\begin{table}[h]
\bc
\begin{tabular}{lccccc}
\hline\hline 
Field & $SU(3)_C$ & $SU(2)_L$ & $U(1)_Y$ & $U(1)_{D}$ & $P_D$\\
\hline
$l_{aL} = \begin{pmatrix}
\nu_{aL}\\
e_{aL}\end{pmatrix}$ & 1 & 2 & $-1/2$ & $0$ & $+$\\
$\nu_{1R}$ & 1 & 1 & 0 & $0$ & $+$\\
$\nu_{2R}$ & 1 & 1 & 0 & $-1$ & $-$\\
$\nu_{3R}$ & 1 & 1 & 0 & $+1$ & $-$\\
$e_{aR}$ & 1 & 1 & $-1$ & $0$ & $+$\\
$q_{aL}= \begin{pmatrix}
u_{aL}\\
d_{aL}\end{pmatrix}$ & 3 & 2 & $1/6$ & $0$ & $+$\\
$u_{aR}$ & 3 & 1 & $2/3$ & $0$ & $+$\\
$d_{aR}$ & 3 & 1 & $-1/3$ & $0$ & $+$\\
$H=\begin{pmatrix}
H^+\\
H^0\end{pmatrix}$ & 1 & 2 & $1/2$ & 0 & $+$\\  
$\eta =\begin{pmatrix}
\eta^0\\
\eta^-\end{pmatrix}$ & 1 & 2 & $-1/2$ & $+1$ & $-$\\  
$\xi$ & 1 & 1 & 0 & $-1$ & $-$\\
$\phi$ & 1 & 1 & 0 & 2 & $+$\\
\hline\hline 
\end{tabular}
\caption[]{\label{tab1} Field representation content of the model.}
\ec
\end{table}    

The Lagrangian contains $\mathcal{L}=\mathcal{L}_{\mathrm{kin}}+\mathcal{L}_{\mathrm{Yuk}}-V$. The kinetic term takes the form, \be \mathcal{L}_{\mathrm{kin}}=\sum_F \bar{F}i\ga^\mu D_\mu F + \sum_S (D^\mu S)^\dagger (D_\mu S)-\fr 1 4\sum_A A_{\mu\nu}A^{\mu\nu},\ee where $F$, $S$, and $A$ run over fermion, scalar, and gauge multiplets, respectively. Additionally, the covariant derivative is defined by \be D_\mu =\pa_\mu + i g_s T_n G_{n\mu} + i g t_j A_{j \mu} + i g' Y B_\mu + i g'' D C_\mu,\ee where $(g_s, g, g', g'')$, $(G_n, A_j, B, C)$, and $(T_n, t_j, Y, D)$ are coupling constants, gauge bosons, and charges according to $(SU(3)_C, SU(2)_L, U(1)_Y, U(1)_D)$ groups, respectively.  

The Yukawa term is obtained by  
\bea \mathcal{L}_{\mathrm{Yuk}} &=& h^d_{ab} \bar{q}_{aL} H d_{bR} + h^u_{ab} \bar{q}_{aL} \tilde{H} u_{bR}+ h^e_{ab} \bar{l}_{aL} H e_{bR} \crn
&&+ h^\nu_{a1}\bar{l}_{aL}\tilde{H}\nu_{1R}+ h^\nu_{a2}\bar{l}_{aL} \eta \nu_{2R}  -\fr 1 2 M_{1}\bar{\nu}^c_{1R}\nu_{1R}-M_{23}\bar{\nu}^c_{2R}\nu_{3R}\crn
&&+y_{12} \bar{\nu}^c_{1R}\nu_{2R}\xi^*+y_{13} \bar{\nu}^c_{1R}\nu_{3R}\xi +\fr 1 2 y_{22}\bar{\nu}^c_{2R}\nu_{2R}\phi+\fr 1 2 y_{33}\bar{\nu}^c_{3R}\nu_{3R}\phi^*+H.c.,\eea
where $\tilde{H}=i\sigma_2H^*$, as usual, $h$'s and $y$'s are dimensionless, while $M$'s have a mass dimension. 

Finally, the scalar potential is given by
\bea V &=& \mu_1^2H^\dag H + \mu_2^2\phi^* \phi + \mu_3^2\eta^\dag \eta + \mu_4^2\xi^* \xi+ \la_1(H^\dag H)^2 + \la_2(\phi^* \phi)^2 + \la_3(\eta^\dag \eta)^2 + \la_4(\xi^* \xi)^2 \crn
&& + \la_5 (H^\dag H)(\phi^* \phi)  + \la_6(\eta^\dag \eta)(\xi^* \xi) + \la_7(H^\dag H)(\eta^\dag \eta)+\la_8(\phi^* \phi)(\eta^\dag \eta)\crn
&& + \la_9(H^\dag H)(\xi^* \xi)+\la_{10}(\phi^* \phi)(\xi^* \xi) + \la_{11} (H^\dag \eta)(\eta^\dag H) + (\kappa_1 H\eta\xi+\kappa_2\xi\xi \phi+H.c.), \eea 
where $\la$’s are dimensionless, while $\mu$’s and $\kappa$'s have a mass dimension. Further, we take $\kappa$'s to be real like other parameters, without loss of generality. 

Necessary conditions for the potential to be bounded from below as well as yielding the desirable vacuum structure are \bea && \mu_{1,2}^2<0,\hs \mu_{3,4}^2>0,\hs \la_{1,2,3,4}>0,\\
&& \la_5>-2\sqrt{\la_1 \la_2},\hs \la_6>-2\sqrt{\la_3 \la_4},\hs \la_8>-2\sqrt{\la_2 \la_3},\\
&& \la_9>-2\sqrt{\la_1 \la_4},\hs \la_{10}>-2\sqrt{\la_2 \la_4},\hs \la_7+\la_{11} \theta(-\la_{11})>-2\sqrt{\la_1 \la_3},\eea where $\theta(x)$ is the Heaviside step function.\footnote{There remain complicated conditions that ensure the quartic coupling matrix to be copositive responsible for vacuum stability \cite{vcs}. Additionally, physical scalar masses squared are required to be positive, but most of these conditions would be equivalent to the given ones.} With the conditions, scalar multiplets develop vacuum expectation values (VEVs), such as $\langle H\rangle= (0,
v/\sqrt2)$, $\langle\phi\rangle=\La/\sqrt2$, $\langle \eta\rangle =0$, and $\langle \xi\rangle =0$, where
\be v^2=\fr{2(\la_5 \mu^2_2-2\la_2 \mu^2_1)}{4\la_1\la_2-\la^2_5},\hs \La^2=\fr{2(\la_5 \mu^2_1-2\la_1 \mu^2_2)}{4\la_1\la_2-\la^2_5},\ee given by potential minimization. We demand $\La\gg v=246$ GeV for consistency with the standard model. The scheme of symmetry breaking is
\bc \begin{tabular}{c} 
$SU(3)_C\otimes SU(2)_L\otimes U(1)_Y\otimes U(1)_D$\\
$\downarrow \La $\\
$SU(3)_C\otimes SU(2)_L\otimes U(1)_Y\otimes P_D$\\
$\downarrow v $
\\ 
$SU(3)_C\otimes U(1)_Q\otimes P_D$\end{tabular} \ec 

Notice that $P_D$ is a residual symmetry of $U(1)_D$ taking the form $P_D=e^{i \al D}$, where $\al$ is a transformation parameter. $P_D$ conserves the vacuum $\La$, thus $P_D \La =e^{i\al 2}\La=\La$, or $\al=k \pi$ for $k$ integer. Hence, we have $P_D=(-1)^{kD}=\{1,(-1)^D\}\cong Z_2$. In other words, the residual symmetry is generated by a dark parity, redefined by \be P_D = (-1)^D.\ee The dark parity of all fields is summarized in the last column of Table \ref{tab1} too. All standard model particles, $\nu_{1R}$, and $\phi$ are even, while $\nu_{2,3R}$, $\eta$, and $\xi$ are odd. It is noteworthy that $\eta$ and $\xi$ have zero VEV due to dark parity conservation.

Expand $H^0=(v+S+i A)/\sqrt{2}$ and $\phi=(\La +S'+iA')/\sqrt{2}$. We find $P_D$-even physical scalars, 
\be H=\begin{pmatrix} G^+_W\\
\fr{1}{\sqrt{2}}(v+c_\varphi H_1+s_\varphi H_2+ i G_Z)
\end{pmatrix},\hs \phi=\fr{1}{\sqrt{2}}(\La -s_\varphi H_1+c_\varphi H_2 + i G_{Z'}).\ee Massless Goldstones $G^+_W\equiv H^+$, $G_Z\equiv A$, and $G_{Z'}\equiv A'$ correspond to gauge bosons $W^+$, $Z$, and $Z'$, respectively. Higgs bosons $H_1\equiv c_\varphi S - s_\varphi S'$ and $H_2\equiv s_\varphi S +c_\varphi S'$ are identical to usual and new Higgs bosons. The $S$-$S'$ mixing angle obeys $t_{2\varphi}\simeq (\la_5 v)/(\la_2 \La) \ll 1$, while the Higgs masses approximate as $m^2_{H_{1}}\simeq (2\la_1-\la^2_5/2\la_2)v^2$ and $m^2_{H_{2}}\simeq 2\la_2 \La^2$.      
        
Expand $\eta^0=(R+i I)/\sqrt{2}$ and $\xi=(R'+i I')/\sqrt{2}$. Further, label $M^2_\eta\equiv \mu^2_3+\fr{\la_7}{2}v^2+\fr{\la_{8}}{2}\La^2$ and $M^2_\xi\equiv \mu^2_4+\fr{\la_9}{2}v^2+\fr{\la_{10}}{2}\La^2$. Charged dark scalar $\eta^\pm$ is a physical field by itself with mass $m^2_{\eta^\pm}=M^2_\eta +\fr{\la_{11}}{2}v^2$. By contrast, neutral dark scalars $R,R'$ and $I,I'$ mix in each pair, defined by mixing angles $\theta_R$ and $\theta_I$ respectively, obeying \be t_{2 R}=\fr{-\sqrt{2}\kappa_1 v}{M^2_{\xi}+\sqrt{2}\kappa_2 \La - M^2_\eta},\hs t_{2 I}=\fr{\sqrt{2}\kappa_1 v }{M^2_{\xi}-\sqrt{2}\kappa_2 \La - M^2_\eta}.\ee
That said, we obtain $P_D$-odd physical fields $R_1=c_{R} R -s_{R} R'$, $R_2=s_{R} R +c_{R} R'$, $I_1=c_{I} I -s_{I} I'$, and $I_2=s_{I} I +c_{I} I'$, with respective masses,
\bea && m^2_{R_{1}}\simeq M^2_\eta + \fr{\kappa^2_1 v^2/2}{M^2_\eta - M^2_\xi-\sqrt{2}\kappa_2 \La},\hs m^2_{R_{2}}\simeq M^2_\xi +\sqrt{2} \kappa_2 \La + \fr{\kappa^2_1 v^2/2}{M^2_\xi +\sqrt{2} \kappa_2 \La - M^2_\eta},\\
&& m^2_{I_{1}}\simeq M^2_\eta + \fr{\kappa^2_1 v^2/2}{M^2_\eta - M^2_\xi +\sqrt{2}\kappa_2 \La},\hs m^2_{I_{2}}\simeq M^2_\xi - \sqrt{2} \kappa_2 \La + \fr{\kappa^2_1 v^2/2}{M^2_\xi -\sqrt{2} \kappa_2 \La - M^2_\eta},\eea where the approximations are due to $|\theta_{R,I}|\ll 1$. 

It is noted that the soft-terms $\kappa_{1,2}$ are not suppressed by any existing symmetry, possibly being as large as the highest scale, i.e. $\kappa_1\sim \kappa_2\sim \La$. However, the $R_1,I_1$ mass splitting, $(m_{R_1}-m_{I_1})/(m_{R_1}+m_{I_1})\sim v /\La \ll 1$, is suppressed as the mixing angles $\theta_{R,I}\sim v/\La\ll 1$ are.

\section{\label{neutrino}Neutrino mass}

It is clear from the Yukawa term that charged leptons, up quarks, and down quarks obtain appropriate masses similar to the standard model. 

\begin{figure}[h]
\bc
\includegraphics[scale=1]{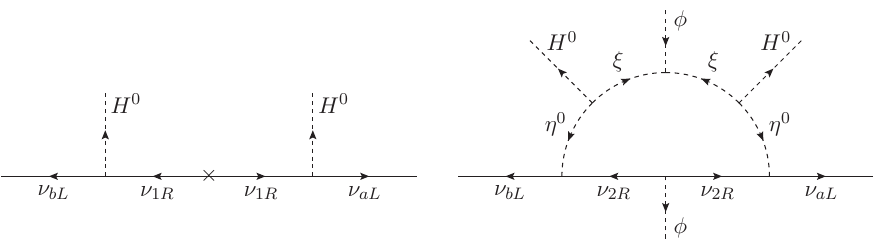}
\caption[]{\label{fig1}Scotoseesaw neutrino mass generation governed by the dark symmetry for right-handed neutrinos.}
\ec
\end{figure}

Concerning neutrinos, the Yukawa term contains the following tree-level masses,
\be \mathcal{L}\supset -\fr 1 2 \begin{pmatrix}
\bar{\nu}_{aL} & \bar{\nu}^c_{1R}
\end{pmatrix}\begin{pmatrix}
0 & m_{a1} \\ m_{b1} & M_{1}
\end{pmatrix}\begin{pmatrix}
\nu^c_{bL} \\ \nu_{1R}
\end{pmatrix}-\fr 1 2 \begin{pmatrix}
\bar{\nu}^c_{2R} & \bar{\nu}^c_{3R}
\end{pmatrix}\begin{pmatrix}
M_{22} & M_{23} \\ M_{23} & M_{33}
\end{pmatrix}\begin{pmatrix}
\nu_{2R} \\ \nu_{3R}
\end{pmatrix}+H.c., \ee
where $m_{a1}=- h^\nu_{a1} v/\sqrt2$, $M_{22}=-y_{22}\La/\sqrt2$, and $M_{33}=-y_{33}\La/\sqrt2$. The first mass matrix takes a canonical seesaw form assuming $M_{1}\gg v$, whereas the second mass matrix has a general form. They are all diagonalized, leading to
\be \mathcal{L}\supset -\fr 1 2 (m_\nu)^{\mathrm{tree}}_{ab}\bar{\nu}_{aL}\nu^c_{bL}-\fr 1 2 M_1 \bar{\nu}^c_{1R}\nu_{1R} -\fr 1 2 M_2 \bar{N}^c_{2R}N_{2R}-\fr 1 2 M_3 \bar{N}^c_{3R} N_{3R}+H.c.\ee The seesaw mechanism yields a small mass,
\be (m_\nu)^{\mathrm{tree}}_{ab}\simeq -\fr{v^2}{2M_{1}} h^\nu_{a1} h^\nu_{1b}, \ee
as illustrated by the left diagram in Fig. \ref{fig1}, while the field $\nu_{1R}$ is heavy with mass $M_{1}$ and decoupled. The second matrix yields two physical states $N_{2,3R}$ related to $\nu_{2,3R}$ through a $2\times 2$ rotation matrix $U$, such as $\nu_{iR} = U_{ij} N_{jR}$ for $i,j=2,3$, with respective masses $M_{2,3}$. 

It is clear that $(m_\nu)^{\mathrm{tree}}_{ab}$ has rank 1, yielding only a massive neutrino $\sim m_{a1}\nu_{aL}$ inappropriate to experiment \cite{pdg}. Fortunately, this tree-level mass matrix receives an one-loop radiative contribution by $P_D$-odd fields, say $\nu_{2R}$, $\eta^0$, and $\xi$, illustrated by the right diagram in Fig. \ref{fig1}. In the mass basis, this diagram is mediated by fermions $N_{2,3}$ and scalars $R_{1,2},I_{1,2}$, governed by the Lagrangian 
\be \mathcal{L}\supset \frac{h^\nu_{a2} U_{2j}}{\sqrt2}\bar{\nu}_{aL}(c_R R_1 +s_R R_2+ic_I I_1 +i s_I I_2 ) N_{jR}+ H.c. \ee Hence, the radiative neutrino mass is computed as
\bea (m_\nu)^{\mathrm{rad}}_{ab} = \frac{h^\nu_{a2}h^\nu_{b2}U^2_{2j}M_j}{32\pi^2}\left(\frac{c^2_Rm^2_{R_1}\ln\frac{M^2_j}{m^2_{R_1}}}{M^2_j-m^2_{R_1}}-\frac{c^2_I m^2_{I_1}\ln\frac{M^2_j}{m^2_{I_1}}}{M^2_j-m^2_{I_1}}+\frac{s^2_R m^2_{R_2}\ln\frac{M^2_j}{m^2_{R_2}}}{M^2_j-m^2_{R_2}}-\frac{s^2_Im^2_{I_2}\ln\frac{M^2_j}{m^2_{I_2}}}{M^2_j-m^2_{I_2}}\right).\eea Because of $s^2_{R,I}\sim v^2/\La^2$ and $(m^2_{R_1}-m^2_{I_1})/(m^2_{R_1}+m^2_{I_1})\sim v^2/\La^2$, the neutrino mass is proportional to $(m_\nu)^{\mathrm{rad}}_{ab} \sim \frac{(h^\nu U)^2 v^2}{32\pi^2 \La}$, which is smaller than the tree-level value by a loop factor, $1/16\pi^2$. This matches the mass hierarchy of the solar and atmospheric neutrinos.   

It is specially stressed that $\xi$ is a mediator field relevant to neutrino mass generation. If it belongs to a more fundamental theory, obtaining a mass much bigger than those of other dark fields, say $\mu_4\gg \mu_3,\La$, thus $M_\xi\simeq \mu_4$. Hence, the right diagram in Fig. \ref{fig1} reduces to that in the minimal scotogenic scheme, where a vertex of the form, $\fr 1 2 \la(H\eta)^2+H.c.$, is identified by $\la=\sqrt2 \kappa_1^2\kappa_2\La/\mu_4^4$. In this case, only the $R_1,I_1$ mass splitting, i.e. $m^2_{R_1}-m^2_{I_1}\simeq \la v^2$, governs the neutrino mass,
\be (m_\nu)^{\mathrm{rad}}_{ab}\simeq \frac{\la v^2}{32\pi^2}\frac{h^\nu_{a2}h^\nu_{b2}U^2_{2j}M_j}{m^2_\eta-M^2_2}\left(1-\frac{M^2_j}{m^2_\eta-M^2_j}\ln\frac{m^2_\eta}{M^2_j}\right),\ee
where $m^2_\eta \equiv (m^2_{R_1}+m^2_{I_1})/2$. We estimate the neutrino mass $(m_\nu)^{\mathrm{rad}}_{ab} \sim \frac{\la (h^\nu U)^2 v^2}{32\pi^2 \La}$, which is more suppressed than the previous case by a factor $\la\sim \La/\mu_4\ll 1$, even given that $\kappa_{1,2}$ are as large as $\mu_4$. In other words, the minimal scotogenic result is much suppressed than the canonical seesaw one, given that $h_{a1}\sim h_{a2}$ and $M_1\sim M_{2,3}$, which does not solve the mass hierarchy of the solar and atmospheric neutrinos. The previous case with $\xi$ and $\eta$ at the same scale is more natural.  

\section{\label{darkmatter}Dark matter}

The model experiences two kinds of dark matter, either a dark Majorana fermion ($N_{2,3R}$) or a dark Higgs boson ($R_{1,2},I_{1,2}$), depending on parameter space that sets which candidate is the lightest of dark fields. They interact with normal fields through $Z'$ and $H_{1,2}$ portals, besides $t$-channels by dark fields. It is noted that the kinetic mixing \cite{kimix} between $U(1)_D$ and $U(1)_Y$ is insignificant, manifestly omitted in this work.  

\subsection{Fermion dark matter}

Let $N_{2R}$ be the lightest of dark fields. It is a dark matter candidate, stabilized by dark parity conservation. $N_{2R}$ can annihilate to normal fields via $s$-channel diagrams by $Z'$ and $H_{2}$ exchanges or $t$-channel diagrams by $\eta^\pm$ and $R_1,I_1$ exchanges. Since $N_{2R}$ is a Majorana fermion, the contribution of $Z'$ to the annihilation cross-section is helicity-suppressed. Additionally, since the dark fields including $N_{2R}$ are at TeV scale, the contribution of $t$-channel diagrams by dark fields is negligible too (cf. \cite{dpsearch}). That said, the $H_2$ portal is crucial to set $N_{2R}$ relic abundance, where $N_{2R}$ dominantly annihilates to top quark, mediated by $H_2$. The annihilation cross-section is computed as  
\be \langle \sigma v_{\mathrm{rel}}\rangle_{N_{2R}}
\simeq \fr{s^2_\varphi m^2_t M^4_2}{8\pi v^2 \La^2(4M^2_2-m^2_{H_2})^2},\ee where we set $M_{23}=0$ for simplicity, thus the dark matter coupling $y_{22}=-\sqrt{2}M_2/\La$, as well as the top coupling $h_t=-\sqrt{2} m_t /v$, have been used. The usual-new Higgs mixing would shift the well-measured standard model Higgs couplings, which constrains $s_\varphi \sim 10^{-2}$ \cite{pdg}. Additionally, taking $m_t=173$ GeV, $v=246$ GeV, and $v/\La=0.1$, the relic density $\Om_{N_{2R}} h^2\simeq 0.1\ \mathrm{pb}/\langle \sigma v_{\mathrm{rel}}\rangle_{N_{2R}}$ is overpopulated for $M_2\neq \fr 1 2 m_{H_2}$. Only if $M_{2}\simeq \fr 1 2 m_{H_2}$, i.e. at the $H_2$ resonance, the dark matter relic density is satisfied, i.e. $\Om_{N_{2R}} h^2\leq 0.12$ \cite{pdg}.   
 
Although the standard model $H_1$ portal does not govern the $N_{2R}$ relic density (which is necessarily set by $H_2$ resonance), it definitely sets the spin-independent (SI) cross-section that determines scattering of $N_{2R}$ with nuclei target in direct detection, because the $H_2$ contribution to effective coupling at quark level $\mathcal{L}_{\mathrm{eff}}\supset \al^S_{q}(\bar{N}^c_{2R}N_{2R})(\bar{q}q)$ is suppressed by $m^2_{H_1}/m^2_{H_2}\ll 1$ compared to that by $H_1$. Here, the $H_1$ contribution is   
\be \al^S_q=-\fr{s_\varphi c_\varphi m_q M_2}{v \La m^2_{H_1}}.\ee The scattering cross-section of dark matter with a nucleon $N=p,n$ is given by \cite{dmdd1} 
\be \sigma^{\mathrm{SI}}_{N_{2R}-N}\simeq \fr{0.49 m^4_N s^2_\varphi M^2_2}{\pi v^2 \La^2 m^4_{H_1}}\simeq \left(\fr{s_\varphi}{10^{-2}}\right)^2\left(\fr{M_2}{\mathrm{TeV}}\right)^2\times 0.68\times 10^{-46}\ \mathrm{cm}^2,\ee which has used $m_N=1$ GeV and $m_{H_1}=125$ GeV, in addition to $v/\La=0.1$. This prediction agrees with the latest measurement $\sigma^{\mathrm{SI}}_{N_{2R}-N}\sim 10^{-46}\ \mathrm{cm}^2$, given that $M_2\sim$ TeV and $s_\varphi\sim 10^{-2}$~\cite{lzexp}.      

\subsection{Scalar dark matter}

As specified, all $R_{1,2}$ and $I_{1,2}$ significantly contribute to radiative neutrino mass, in addition to $N_{2,3R}$. Additionally, since $\theta_{R,I}$ are small, one can identify $\eta^0\simeq R_1, I_1$, while $\xi\simeq R_2,I_2$. Although $N_{2,3}$ and $\eta^0\simeq R_1,I_1$ have features similar to those in the minimal scotogenic setup and the previous setup \cite{dongloi}, the existence of $\xi\simeq R_2,I_2$ that contributes to neutrino mass is novel and worth exploring. In what follows, we suppose $R_2\simeq \sqrt{2}\Re(\xi)$ to be the lightest of dark fields, providing a dark matter candidate. The nature of dark matter is just a scalar scoto-singlet, where ``scoto'' means darkness. This candidate communicates with normal matter via $H_{1,2}$ portals or interacts with Higgs $H_{1,2}$ via $R_1\simeq \sqrt{2}\Re(\eta^0)$ dark field.  

Since $R_2$ is heavy at TeV scale, in the early universe $R_2$ dominantly annihilates to top quarks via $s$-channel $H_{2}$ portal, which sets its relic abundance by $H_2$ resonance, $m_{R_2}\simeq \fr 1 2 m_{H_2}$, such that $\Om_{R_2} h^2 \simeq 0.1\ \mathrm{pb}/\langle \sigma v_{\mathrm{rel}}\rangle_{R_2}$, where
\be \langle \sigma v_{\mathrm{rel}}\rangle_{R_2} \simeq \fr{s^2_\varphi m^2_t (\sqrt{2} \kappa_2+\la_{10} \La)^2}{8 \pi v^2(4 m^2_{R_2}-m^2_{H_2})^2}. \ee It is noted that the contributions of $H_1$ and $R_1$ portals are small, as omitted. 

Concerning direct detection, $R_2$ scatters with nuclei target via $H_1$ portal, where notice that the $H_2$ portal negligibly contributes, while the $Z/Z'$ portal is suppressed due to a $R_2,I_2$ mass splitting. At quark level, the relevant effective interaction is $\mathcal{L}_{\mathrm{eff}}\supset 2\la_q m_{R_2} R^2_2\bar{q}q$, where \be \la_q = \fr{\la' m_q}{2 m_{R_2}m^2_{H_1}},\hs \la'=\la_9-\fr{\sqrt{2}s_\varphi \kappa_2}{v}, \ee where $\la_q$ is the effective coupling related to the $R_2$-$H_1$ coupling $\la'$ in $V\supset \fr 1 2 \la' R^2_2 H_1$. It is noted that the scalar candidate has only spin-dependent and even interactions with quarks, while interactions with gluons are loop-induced and small. 

The $R_2$-nucleon ($N=p,n$) scattering amplitude is obtained by summing over quark-level contributions multiplied with relevant nucleon form factors. Therefore, the $R_2$-$N$ scattering cross-section is derived as follows
\be \sigma_{R_2-N} \simeq  \fr{4 m^2_N}{\pi} \la^2_N, \ee where $\la_N/m_N \simeq 0.35 \la'/(2m_{R_2} m^2_{H_1})$, which summarizes quark contributions \cite{dmdd2}. This yields
\be \sigma_{R_2-N} \simeq \left(\fr{\mathrm{TeV}}{m_{R_2}}\right)^2\left(\fr{\la'}{0.04}\right)^2\times 10^{-46}\ \mathrm{cm}^2,\ee in good agreement with the latest measurement, namely  $\sigma_{R_{2}-N}\sim 10^{-46}\ \mathrm{cm}^2$, given that $m_{R_2}\sim$~TeV and $\la'\sim 0.04$ \cite{lzexp}. With the help of $s_\varphi\simeq \fr 1 2 (\la_5/\la_2)(v/\La)$, the last condition translates to $\la'\simeq \la_9-(1/\sqrt{2})(\la_5/\la_2)(\kappa_2/\La) \sim 0.04$, making a constraint on $\la_9$ and $\la_5/\la_2$ as appropriate, for the soft-term $\kappa_2\sim \La$ as already fixed.  

\section{\label{bound} Collider bounds}

Our dark fields including dark matter candidate are heavy at TeV. Thus, they may present compelling signals given in form of large missing energy at the LHC, which are worth exploring.  

\subsection{Higgs portals: signal suppressed}

The LHC studies novel visible signals recoiled against large missing energy carried by a pair of dark matter, possibly making significant constraints on interactions between dark matter and quarks exchanged by a neutral Higgs boson, such as $H_1, H_2$. In this model, both dark matter candidates $N_2$ and $R_2$ are set by $H_2$ mass resonance, $m_{\mathrm{DM}}\simeq \fr 1 2 m_{H_2}$. Additionally, all the dark fields are radically heavier than the usual $H_1$ field. 

Given that $H_1$ is produced at the LHC, it cannot decay to a pair of dark fields as kinematically forbidden. Alternatively, when $H_2$ is produced at the LHC, its decay to a dark matter candidate either $N_2$ or $R_2$ is strongly suppressed by a phase space factor $1-4m^2_{\mathrm{DM}}/m^2_{H_2}$, which is associated with the decay width $\Ga(H_2\to \mathrm{DM}+\mathrm{DM})$.

In this point of view, there is no missing energy stored in dark matter product associated with $H_{1,2}$, which govern dark matter direct detection and relic density, respectively.     

\subsection{Gauge portals: dilepton signal}

We assume $\eta^\pm$ is lighter than every dark field, except for dark matter $N_2$. Hence, $\eta^\pm$  decays only to $N_2$ associated with an effective charged lepton, i.e. $\mathrm{Br}(\eta^-\to l N_2)=\mathrm{Br}(\eta^+\to l^c N_2)=1$. Since the LHC is energetic, a pair of dark scalars $\eta^\pm$ can be created from $pp$ collision, i.e. $pp\to \eta^+\eta^-$, then followed by $\eta^\pm$ decays to $N_2$'s, i.e. $\eta^-\to l N_2$ and $\eta^+ \to l^c N_2$. 

The LHC looks for a dilepton signal $ll^c$, recoiled against large missing energy ($E\!\!\!\!/_T$), carried by a pair of $N_2N_2$. The dilepton cross section is given by 
\bea \sigma(pp \to ll^c+E\!\!\!\!/_T)&=&\sigma(pp\to \eta^+\eta^-\to ll^c N_2 N_2)\crn
&=&\sigma(pp \to \eta^+\eta^-)\times \mathrm{Br}(\eta^-\to l N_2)\times \mathrm{Br}(\eta^+\to l^c N_2), \eea by the narrow width approximation, where the last two branching ratios equal 1.  

The process $pp \to \eta^+\eta^-$ is dominantly contributed by $s$-channel $\ga,Z$ exchanges. The relevant cross-section is given at quark level as 
\be \sigma(qq^c\to \eta^+\eta^-)\simeq \fr{\pi \al^2}{36E^2}\left(1-\fr{m^2_{\eta^\pm}}{E^2}\right)^{3/2} \left[Q^2_q +\left(\fr 1 2 -s^2_W\right) \fr{Q_q v_q }{s^2_W c^2_W}+\left(\fr 1 2 -s^2_W\right)^2\fr{v^2_q+a^2_q}{s^4_W c^4_W} \right],\ee where $v_q=T_{3q}-2s^2_W Q_q$ and $a_q=T_{3q}$ denote the $Z$-quark couplings, as usual, while the incident quark energy obeys $E=\fr 1 2 \sqrt{s}>m_{\eta^\pm}\gg m_Z$. 

The dark scalar $\eta$ has the same statistic and quantum numbers with well-studied, left-handed slepton in SUSY, which matches $\sigma(qq^c\to \eta^+\eta^-) = \sigma(qq^c\to \tilde{l}\tilde{l}^*)$. The LHC \cite{ppvv1} examined slepton-pair production, then decayed to a dilepton signal plus missing energy, $pp \to \tilde{l}\tilde{l}^*\to ll^c \tilde{\chi}^0_1\tilde{\chi}^0_1$, with $\mathrm{Br}(\tilde{l}\to l \tilde{\chi}^0_1)=1$, bounding charged slepton mass to be $m_{\tilde{l}}>700$ GeV. Hence, the SUSY result can apply to our case without change, \be m_{\eta^\pm}>700\ \mathrm{GeV}.\ee

\section{\label{flvg2} Lepton flavor violation and muon $g-2$}

In the standard model, lepton flavors are separately conserved as far as neutrinos are massless. However, the observed neutrino oscillations reveal a direct evidence for lepton flavor violation. This flavor violation in neutrino propagator hints that it may occur in charged lepton sector, such as $\mu \to e \ga$, for instance. Indeed, any observation of charged lepton flavor violation would be a crucial indication for new physics and enhance our knowledge of lepton sector \cite{revmug2}. 

Concerning charged lepton flavor violation in the present model, the most dangerous process is related to $\mu\to e \ga$ through exchange of dark fields $(\eta^\pm,N_{2,3})$ in one-loop diagrams, governed a fundamental interaction, $\mathcal{L}_{\mathrm{Yuk}}\supset h^\nu_{a 2} \bar{l}_{aL} \eta \nu_{2R} +H.c.$, where $\nu_{2R}=U_{2j}N_{jR}$ for $U^*_{2j}U_{2j}=1$, and the charged flavor $a$ is regarded as a physical state, without loss of generality. It is easily derived, 
\be \mathrm{Br}(\mu \to e\ga) \simeq \fr{\al v^4 |h^{\nu*}_{\mu 2} h^\nu_{e 2 }|^2}{48 \pi m^4_{\eta^\pm}}\simeq 1.89\times 10^{-13}\left(\fr{1\ \mathrm{TeV}}{m_{\eta^\pm}}\right)^4\left(\fr{|h^{\nu*}_{\mu 2} h^\nu_{e 2 }|}{10^{-3}}\right)^2.\ee The current sensitivity reported by MEG experiment bounds $\mathrm{Br}(\mu \to e \ga)\simeq 4.2 \times 10^{-13}$ \cite{MEG}, leading to $|h^{\nu*}_{\mu 2} h^\nu_{e 2 }| \sim 10^{-3}$ and $m_{\eta^\pm}\sim 1$ TeV.

\begin{figure}[h]
\bc
\includegraphics[scale=1]{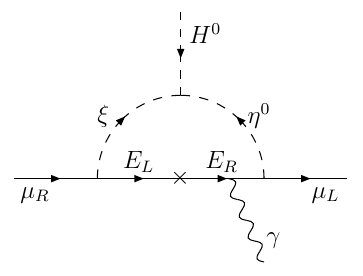}
\caption[]{\label{fig2} Muon $g-2$ anomaly induced by dark fields.}
\ec
\end{figure}

It is easily shown that $\eta^\pm,N_{2,3}$ contribute to the muon anomalous magnetic moment $(g-2)$ via a loop diagram similar to $\mu\to e \ga$. However, this new contribution is not significant and negative. Hence, the present model does not explain the muon $g-2$ anomaly \cite{mug2}. We think of a gauge completion for the current theory in general and right-handed neutrinos in particular. Since usual charged leptons are associated with usual neutrinos, there might exist exotic charged leptons that are coupled to dark right-handed neutrinos. In other words, there is a dark mirror of charged leptons, called $E_a\sim (1,1,-1,1)$, similar to neutrino sector.\footnote{Here, the quantum numbers of $E_a$ are given under the gauge symmetry $SU(3)_C\otimes SU(2)_L\otimes U(1)_Y \otimes U(1)_D$, respectively, in similarity to those in Table \ref{tab1}, and notice that $E_a$ possess $P_D=-1$.} In such case, fundamental interactions arise, as given by \be \mathcal{L}\supset h_{ab}\bar{l}_{aL}\tilde{\eta}E_{bR}+f_{ab} \bar{E}_{aL} e_{bR}\xi^* - (m_E)_{ab} \bar{E}_{aL} E_{bR}+H.c.\ee The muon $g-2$ anomaly is manifestly solved by radiative contributions of $\eta^0\simeq R_1, I_1$, $\xi\simeq R_2,I_2$, and $E$'s as depicted in Fig. \ref{fig2}.

Indeed, let the usual charged flavor ($e_a$) be a physics state, as before. Additionally, let the exotic charged flavor ($E_a$) be a physical state by itself, without loss of generality. The new physics contribution to muon $g-2$ is computed as   
\bea \Delta a_\mu &=& \sum_{E}\fr{h_{\mu E} f_{E\mu} }{16\pi^2}\left(\fr{m_E}{m_\mu}\right)
\left[s_R F\left(\fr{m^2_E}{m^2_\mu},\fr{m^2_{R_1}}{m^2_\mu}\right)-s_I F\left(\fr{m^2_E}{m^2_\mu},\fr{m^2_{I_1}}{m^2_\mu}\right)\right.\crn
&& \left. -s_R F\left(\fr{m^2_E}{m^2_\mu},\fr{m^2_{R_2}}{m^2_\mu}\right)+s_I F\left(\fr{m^2_E}{m^2_\mu},\fr{m^2_{I_2}}{m^2_\mu}\right)\right],\eea where the sum is taken over physical states $E=\{E_1,E_2,E_3\}$, and the loop function is given by 
\be F(\al,\beta)=\int_0^1 d x \fr{x^2}{x^2+(\al-1)x + \beta (1-x)}.\ee Above, $R_1,I_1$ and $R_2,I_2$ mass splittings also contribute to the muon $g-2$, but it is even nonzero for degenerate masses, opposite to \cite{babu}.

It is clear that $s_R\sim s_I\sim v/\La\sim 0.1$ and $F(\al,\beta)\sim (\al,\beta)^{-1}$ for $\al,\beta\gg 1$. Taking $m_{R_{1,2}},m_{I_{1,2}}\sim m_E$, we derive the order of magnitude 
\be \Delta a_\mu \sim 10^{-9}\times \left(\fr{h_{\mu E}f_{E\mu}}{10^{-2}}\right)\left(\fr{700\ \mathrm{GeV}}{m_E}\right),\ee in agreement with the deviation recently measured, \be \Delta a_\mu =a_\mu(\mathrm{Exp})-a_\mu(\mathrm{SM}) = (251\pm 59)\times 10^{-11}\sim 10^{-9},\ee at $4.2\sigma$ from the standard model prediction \cite{mug2}. Here, $E$ must have a mass larger than that of dark matter either $R_2$ or $N_2$, as desirable, while $f_{\mu E}h_{E\mu}\sim 10^{-2}$.

In this view, the existence of $\xi$ is crucial to address the muon $g-2$ anomaly, in addition to neutrino mass generation and dark matter stability, as above given.
     
\section{\label{concl}Conclusion}

We have shown that the right-handed neutrinos---the counterparts of usual left-handed neutrinos---may be fundamental building-blocks of a dark force. They provide the need for a minimal scotoseesaw mechanism that explains simultaneously neutrino masses and dark matter stability. 

To solve the mass hierarchy of the solar and atmospheric neutrinos through a loop factor in the scotoseesaw mechanism, it naturally introduces an inert scalar singlet $\xi$. Since, otherwise, the loop-induced neutrino mass is much suppressed than the tree-level seesaw mass.

Although the dark force $Z'$ does govern dark matter observables as it interacts with normal matter and even dark matter so weak, the Higgs portals play the role. Here, the new Higgs $H_2$ resonance sets dark matter relic density, while the usual Higgs $H_1$ dictates dark matter direct detection. 

A novel dilepton signal can be looked for at the LHC, which recoils against large missing energy associated with fermion dark matter. By contrast, if dark matter has a nature of scalar particles, it is hard to search for this kind of candidate.             

The current setup agrees with the constraint from $\mu\to e \gamma$. However, it cannot solve the muon $g-2$ anomaly. Given that this anomaly is confirmed, embedding the current model to a more fundamental theory is indeed natural.    

\section*{Acknowledgement}

This research is funded by Vietnam National Foundation for Science and Technology Development (NAFOSTED) under grant number 103.01-2023.50.

\end{document}